\documentclass{nature}
\usepackage[pdftex]{graphicx}
\usepackage{epstopdf}
\usepackage[pdftex]{graphicx}
\usepackage{lineno}
\usepackage{xcolor}

\title{Revealing multiple classes of stable quantum emitters in hexagonal boron nitride with correlated cathodoluminescence, photoluminescence, and strain mapping 
}
\author{Fariah Hayee$^{1}$*, Leo Yu$^2$, Jingyuan Linda Zhang$^{3}$, Christopher J. Ciccarino$^4$, Minh Nguyen$^5$, Ann F. Marshall$^6$, Igor Aharonovich$^5$, Jelena Vu\v{c}kovi\'{c}
$^{3}$, Prineha Narang$^4$, Tony F.~Heinz$^{2,7}$, Jennifer A. Dionne$^8$*}

\begin{document}
\maketitle
\begin{affiliations}
 \item Department of Electrical Engineering, Stanford University, 450 Serra Mall, Stanford, CA 94305, USA
 \item Department of Applied Physics, Stanford University, Stanford, CA 94305, USA
 \item E.~L.~Ginzton Laboratory, Stanford University, Stanford, CA 94305, USA
 \item John A. Paulson School of Engineering and Applied Sciences, Harvard University, Cambridge, MA 02138, USA
 \item School of Mathematical and Physical Sciences, University of Technology Sydney, Ultimo, NSW, 2007, Australia
 \item Stanford Nano Shared Facilities, Stanford University, Stanford, CA 94305, USA
 \item SLAC National Accelerator Laboratory, Menlo Park, CA 94025 
 \item Department of Materials Science and Engineering, Stanford University, Stanford, CA 94305, USA\\
* e-mail: fariah@stanford.edu; jdionne@stanford.edu
\end{affiliations}
\begin{abstract}
Single photon emitters (SPEs) in solids have emerged as promising candidates for quantum photonic sensing,~\cite{kucsko2013nano-thermo-NVdiamond, taylor2008magnetometer, aslam2017nanoscale} communications,~\cite{wehner2018quantum, liao2017satellite} and computing. ~\cite{awschalom2018quantum, Wrachtrup2018rev} Defects in hexagonal boron nitride (hBN) exhibit high-brightness, room-temperature quantum emission,~\cite{tran2016quantum, jungwirth2016temperature,grosso2017tunable} but their large spectral variability and unknown local structure significantly challenge their technological utility. Here, we directly correlate hBN quantum emission with the material's local strain using a combination of photoluminescence (PL), cathodoluminescence (CL) and nano-beam electron diffraction. Across 40 emitters and 15 samples, we observe zero phonon lines (ZPLs) in PL and CL ranging from 540-720 nm. CL mapping reveals that multiple defects and distinct defect species located within an optically-diffraction-limited region can each contribute to the observed PL spectra. Local strain maps indicate that strain is not required to activate the emitters and is not solely responsible for the observed ZPL spectral range. Instead, four distinct defect classes are responsible for the observed emission range. 
One defect class has ZPLs near 615 nm with predominantly matched CL-PL responses; it is not a strain-tuned version of another defect class with ZPL emission centered at 580 nm. A third defect class at 650 nm has low visible-frequency CL emission; and a fourth defect species centered at 705 nm has a small, $\sim$10 nm shift between its CL and PL peaks. All studied defects are stable upon both electron and optical irradiation. Our results provide an important foundation for atomic-scale optical characterization of color centers, as well as a foundation for engineering defects with precise emission properties.
\end{abstract}


Defects in hexagonal boron nitride (hBN) exhibit promising quantum-optical properties including narrow linewidth,~\cite{jungwirth2016temperature} high brightness,~\cite{tran2016quantum} high photostability,~\cite{sontheimer2017photodynamics, exarhos2017optical} and high emission into the zero phonon line (ZPL).~\cite{martinez2016bulkhBN} However, both intrinsic and engineered defects in mono- and multi-layer films exhibit large spectral variability.~\cite{tran2016ACSNano, choi2016engineering, chejanovsky2016structural} Additionally, some emitters exhibit distinct polarization profiles for absorption and emission~\cite{jungwirth2017optical} as well as different quantum efficiencies for different excitation wavelengths, suggesting a complicated electronic structure.~\cite{schell2018quantum, exarhos2017optical, sontheimer2017photodynamics} Strain has  been postulated to be responsible for such spectral variability, \cite{grosso2017tunable} where density functional theory calculations have predicted multiple defect structures as candidates.~\cite{Reimers:2018, tawfik:2017} However, the origin of the reported spectral variability of single photon emitters still remains unknown. Establishing a  correlation between the emission of a SPE and its local crystallographic environment is a necessary step towards defect identification and application in various quantum technologies. 

Recent progress in super-resolution optical microscopy and cathodoluminescence (CL) spectroscopy has enabled localization of radiative emission beyond the diffraction limit. For example, quantum emission from hBN has been visualized with 10 nm and 80 nm spatial resolution using super-resolution techniques~\cite{feng2018superres} and CL spectroscopy,~\cite{bourrellier2016bright, meuret2015CLBunching} respectively. However, to date, these techniques have not provided a one-to-one correlation between emission and any structural property, including strain and defect structure. In parallel, aberration-corrected electron microscopy~\cite{alem2009atomically, jin2009hBNHRTEM} and scanning tunneling microscopy~\cite{wong2015STM, lin2017STM, barja2018STM} have revealed individual defect structures in mono- and multi-layered materials, but without correlation to the defects' optical signatures. Notably, locating SPEs within an electron microscope can be extremely challenging since most samples posses a small density of visible SPEs. Further, most  transmission electron microscopes (TEMs) lack optical excitation and detection capabilities, and the effect of the electron beam on single photon emitting defects remains unknown. 
Here, we correlate the PL and CL spectroscopic signatures of hBN visible-frequency SPEs with their local strain using scanning transmission electron microscopy (STEM). First, we establish a direct correlation between a SPE's PL and CL emission, showing two classes of stable emitters: those for which both electron-beam excitation and 532-nm laser excitation generate the same emission spectra, and those for which CL peaks are spectrally shifted. We then use hyperspectral STEM-CL mapping and find that multiple radiative defects are localized within a $\sim$50 nm region. Accordingly, defect-defect interactions and/or multiple radiative transitions within the joint density of electronic states can influence the far-field PL spectra. Using high-resolution election imaging, we find that about $30\%$ of the studied emitters are positioned at least 12 nm away from any possible flake edges. 
Then, using nano-beam electron diffraction, we measure strain variations in the $\sim$20nm regions surrounding the SPEs. A number of SPE wavelengths are observed without considerable strain variation, indicating that strain is not solely responsible for the observed spectral variability. Additionally, our results show that emitters can be classified into at least four distinct SPE defect species, with each species responsible for the 580nm, 615nm, 650nm, and 705nm spectral regions.

\begin{figure}[!htbp]
    \centering
    \includegraphics[width = 1\textwidth, angle = 0]{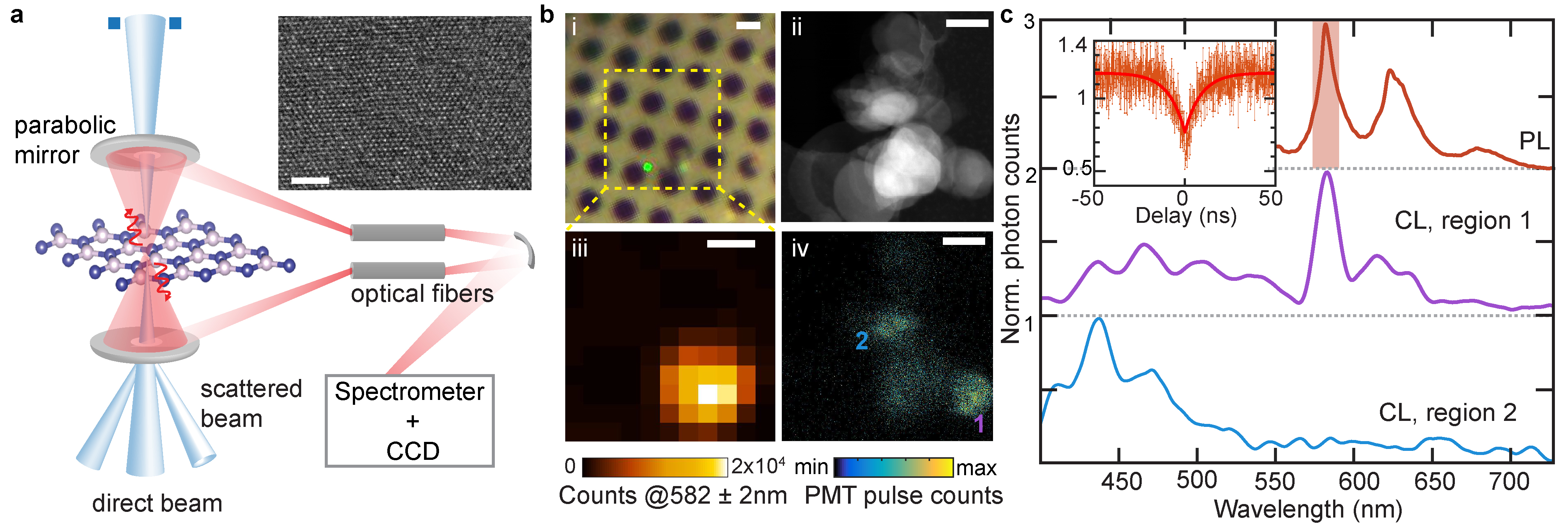}
    \caption{\textbf{Correlated optical and electron characterization  of  quantum  emitters  in  hBN.} 
    (a) Schematic  illustration  of  the cathodoluminescence setup in a TEM. Two parabolic mirrors surround the sample, and the resulting signal is collected via optical fibers and directed to a spectrometer. An alternate light path to a PMT (not shown) can be selected by changing the mirror position. The inset shows a high-resolution TEM image of the sample. Scale bar is 2 nm. (b) Optical image (i) and photoluminescence map for emission at 582 nm (iii); STEM-HAADF image (ii) and panchromatic CL map (iv) of the same hBN flake on a holey-carbon TEM grid.  Scale bars are 1 $\mu$m for the optical images and 200 nm for the STEM images. (c) PL spectrum at the optical hotspot and CL spectra at the regions marked in (b). The corresponding g$^{(2)}$ curve is plotted in the inset. The vertical shaded area in the PL spectrum corresponds to the studied ZPL. The g$^{(2)}$ measurement is not background corrected.}
    \label{fig:experimentation}
\end{figure}

Our experimental STEM-CL setup is depicted in  Fig.~\ref{fig:experimentation}(a). A hBN multi-layer (1-5 layers) nanoflake solution in ethanol and water is drop-cast on a holey carbon TEM substrate for PL, CL and TEM imaging. First, we identify emission centers in multilayer hBN flakes using confocal PL mapping with a 532 nm excitation laser. With such sub-bandgap excitation,~\cite{gil2016hBNbandgap} we excite defects lying deep in the bandgap. We find bright emission ranging between 540-720 nm, with spectral full-widths at half-max (FWHM) ranging from 4 to 12 nm at room temperature. Figure \ref{fig:experimentation}(b) and (c) include an optical micrograph, PL map and PL spectrum of one such representative emitter with a ZPL of 582 nm and two phonon replicas at 622.5 and 680.7 nm (separated by 139 meV and 161 meV from the ZPL). We verify that the emission lines are indeed of single-photon nature by characterizing the second-order autocorrelation function $g^{(2)}(\tau)$ using a Hanbury Brown and Twiss (HBT) setup. Despite the high background contribution at room temperature, the $g^{(2)}(0)$ dip still approaches 0.5, suggesting the non-classical nature of emission. Additional emitters with $g^{(2)}(0)$ well below 0.5 are included in SI; the similarity between our observed PL spectra to those of the reported hBN quantum emitters \cite{tran2016quantum, jungwirth2016temperature, koperski2018optica} further supports the non-classical nature of the emitters we study.

We next identify the same emitters in the TEM (see methods). The process of light generation in CL is incoherent: electrons excite a high energy bulk plasmon ($\sim$20 eV for hBN) which decays within a \emph{fs} time period to generate multiple electon-hole pairs.~\cite{bourrellier2016bright} As shown in Fig.~1(a), we use a focused electron probe (typical diameter of 1.5 nm) and collect the emitted light using two parabolic mirrors positioned above and below the sample. A spectrometer and CCD enable spectroscopic mapping of defects correlated with high-resolution transmission electron microscopy. An alternate light path with a photomultiplier tube (PMT) is used to rapidly map panchromatic emission. A high-angle annular dark field (HAADF) image and corresponding panchromatic-CL (pan-CL) image is shown in Fig.~1(b). The HAADF image shows multiple overlapping nanoflakes, where the thickest nanoflake-stack appears the brightest. As seen from Fig.~1(b) panels iii and iv, within the region of brightest PL, the pan-CL image has two main emission `hotspots'. We collect spectral signatures from each region, with representative smoothed point-spectra shown in Fig.~\ref{fig:experimentation}(c). We find that the CL spectrum of region 1 matches well with the PL spectrum whereas region 2 has a broad spectral peak at 420 nm that is prevalent throughout this flake as well as in many additional flakes. The ZPL in region 1 has a broader FWHM in the CL spectrum (10 nm) than the PL spectrum (6 nm), a likely consequence of the CL spectrometer resolution ($\sim$10 nm). In addition to the ZPL, two other CL peaks are seen at 614 nm and 633 nm; as will be shown later, the first is from another nearby defect while the second is likely a phonon sideband. 

We zoom into region 1 of Figure 1(b)iv and use hyperspectral CL imaging to map local emission with $\sim$15 nm spatial resolution.  Figure 2(a) shows a HAADF image of region 1. CL spectra are collected in 5 nm spatial increments throughout the region denoted by the white dotted box. The spectral images are analyzed with non-negative matrix factorization \cite{hyperspy} (see methods). The resultant three principal CL spectra and their corresponding spatial maps are included in Fig.~2(a). As seen, one constituent spectrum exhibits a 582 nm ZPL, shown by the red spectrum and the red-colored spectral weight map (labelled c$_1$); this is the single-photon emitter whose PL and CL closely match. Two additional emitters are located within 30 and 80 nm of this central emitter, respectively: one emitter is near the top of the mapped region, with a peak emission at 614 nm (purple map, c$_2$), while the other is near the bottom right with a peak emission at 600 nm (blue map, c$_3$). Notably, even though the PL spectrum seems to only show a ZPL and its phonon side-band, the CL spectrum actually shows another emitter at 614 nm near the phonon side-band peak. The centroid of each emitter is boxed on the HAADF image, whose intensity is due to sample thickness variations here. A change of thickness or contrast indicates a flake edge; our analysis (SI S7) indicates that the SPE is at least 17 nm away from any flake edge, where the 600 nm emitter can be situated at an edge. Across 15 different SPEs with clear HAADF image contrast, we find 5 that are at least 12 nm away from any flake edges and 9 within 5 nm of the edges. 

\begin{figure}[!htbp]
    \centering
    \includegraphics[width = .5\textwidth, angle = 0]{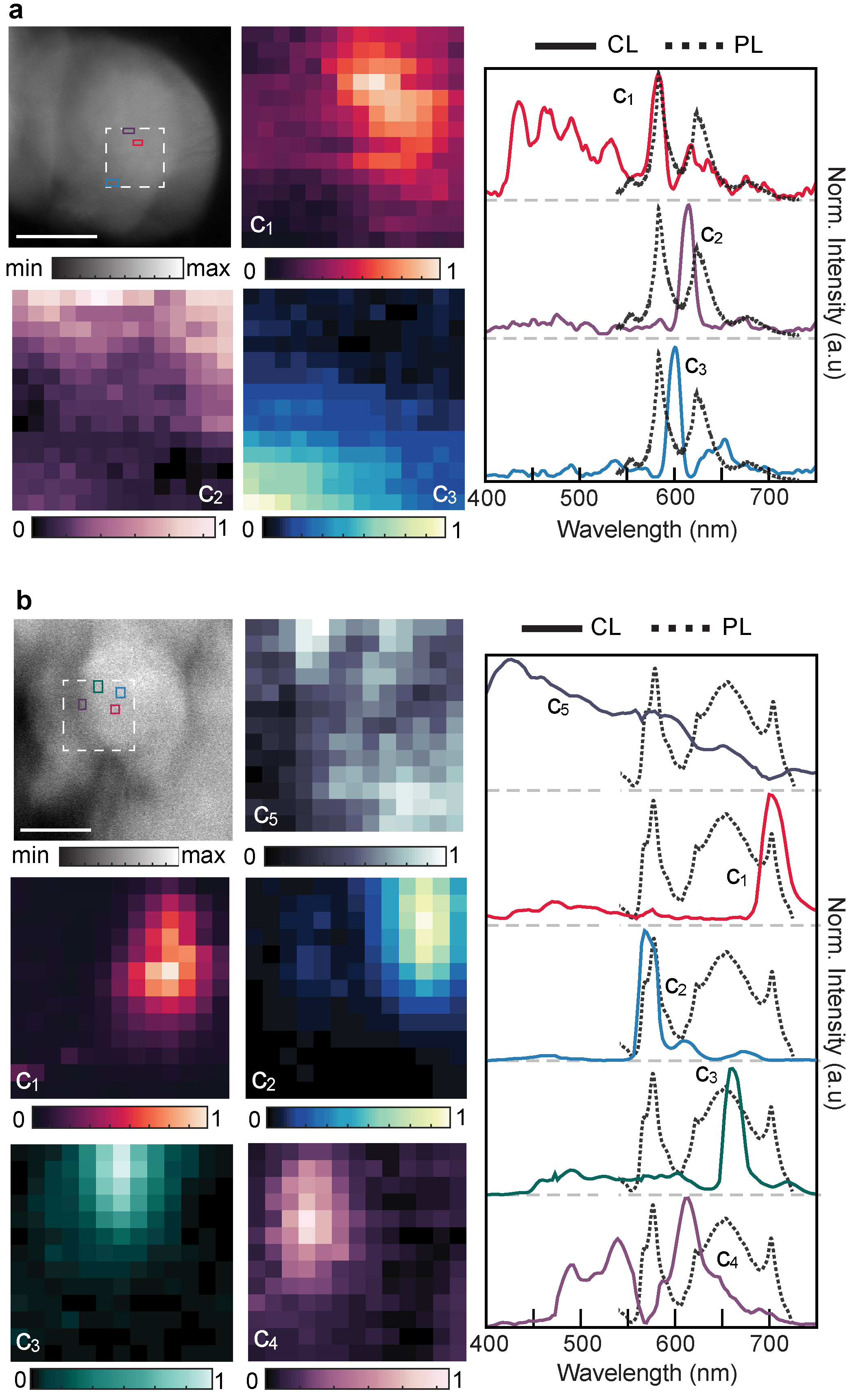}
    \caption{\textbf{Cathodoluminescence mapping reveals multiple and distinct quantum emitter species within a diffraction-limited optical spot} (a) STEM-HAADF image of region 1 in Fig.~1(b) and cathodoluminescence spectral weights (normalized to each component) of the white marked region. The red, purple, and teal colored boxes in the HAADF image are positions of the brightest pixels of the c$_1$, c$_2$ and c$_3$ spectral components. The decomposed spectra are plotted next to the spectral weights. The PL spectrum of the corresponding hotspot is also plotted as a black dotted line. Scale bar in the HAADF image is 100 nm and each pixel in the CL spectral map is 5 nm. (b) CL spectral maps of another region hosting two different quantum emitters at 571 nm and 704 nm. The PL spectrum is a result of multiple defects as seen through CL spectral weight mapping. Pixel size in the CL map is 14.8 nm and the scale bar in the HAADF image is 200 nm.} 
    \label{fig:CL specral mapping}
\end{figure}

The spectrum of Fig.~2a/c$_1$ also illustrates a recurring emission profile with four higher energy peaks at 433, 461, 490 and 532 nm. 
Such CL emission is recurrent in 3 spatially-distinct emitters with PL emission profile consisting of a ZPL (near 578-590 nm) and possible phonon-replica doublet (Fig.~S5). 
In all three cases, the highest two energy peaks (433 and 461 nm) are separated by nearly the hBN phonon energy $\sim$180 meV. Such CL spectra could emerge from closely spaced (sub-20 nm) emitters and their phonon sidebands. The recurrent nature of the higher energy peaks suggests that such emitter(s) at 490 and 532 nm commonly accompany the 580 nm emitter. 

PL spectra can also, in some cases, result from multiple single-photon emitters within one diffraction-limited spot. Figure 2(b) illustrates one such case. As seen, the PL spectrum (black dotted line) exhibits two distinctive, sharp peaks at 571 nm and 704 nm (g$^{(2)}$ in Fig.~S3). Through hyperspectral CL mapping, we see that the two emission lines are actually from two different emitters which are separated by 50 nm, shown in the blue and red CL maps respectively. The smaller PL peak at 625 nm is likely a combination of the 570 nm emitter phonon sideband (blue solid spectrum with a small peak at 620 nm) and an additional radiative defect active only in CL emitting at 625 nm (purple map). A fourth point defect at 682 nm is also evident upon electron beam excitation, shown in the green map. Finally, we observe broad emission peaking at 430 nm, a signal present in most flakes and generally very delocalized. This broad ultraviolet emission can be associated with prevalent impurity atoms like carbon; previously blue PL after ultraviolet excitation has also been linked to carbon centers \cite{katzir1975C-impurity}. Thus, hyperspectral CL mapping allows the delineation of PL spectra to further reveal the existence of multiple optically active defects, masked by a combination of their spatial positions with respect to each other and the diffraction limit.

The spectral maps of Fig.~2 also provide an estimate of the carrier diffusion lengths. Linescans of intensity across particular defect centers are plotted in Fig.~S9. We find that the emission intensity decreases to $1/e$ of its maximum value at 15-60 nm for various emitters, which enables us to distinguish emitters with similar resolution. As discussed in SI, the high density of structural defects and dislocations near emitters may account for such short diffusion lengths.
\begin{figure}[!htbp]
    \includegraphics[width = 0.5\textwidth, angle = 0]{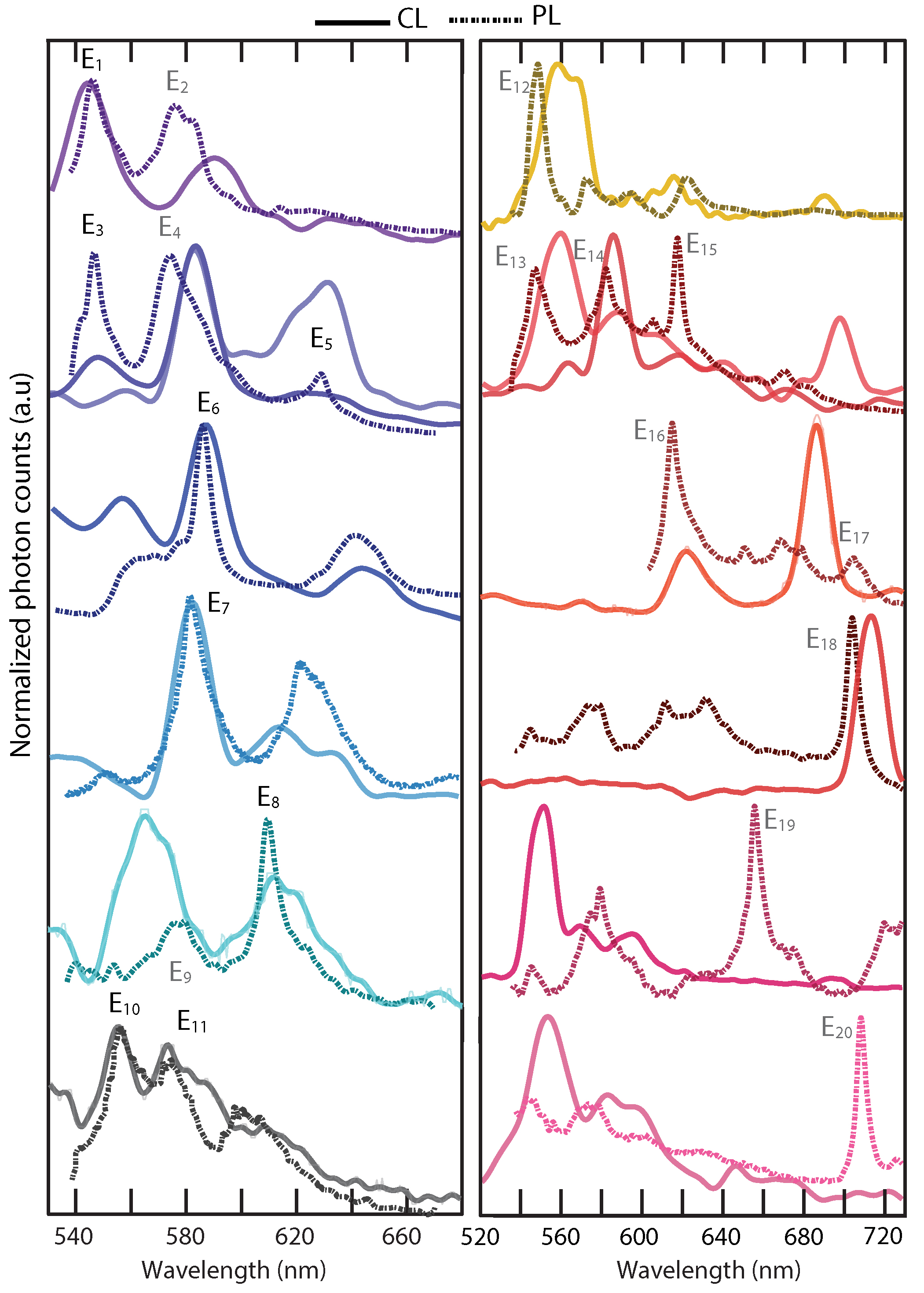} 
    \centering
    \caption{\textbf{Correlating stable emitter PL and CL reveals two types of behavior.} Most of the left column emitters (names in black font) have closely-matched ZPLs upon optical and electron beam excitation. Emitters on the right side and $\mathrm{E_2}$ and $\mathrm{E_4}$ have variable spectral shifts between their PL and CL ZPLs.
    }
    \label{fig:CL_PL}
\end{figure}

Using our spectral mapping technique, we build a library of PL-CL signatures for 40 stable emitters across 15 different samples. Fig.~\ref{fig:CL_PL} presents examples of the PL and CL responses, whose PL spectra are reproducible after electron beam imaging. Two distinct behaviors are observed: one class of emitters exhibits PL emission lines that are well-matched to CL emission lines, while the second class has a shifted CL signature from the PL; of this second class, some emitters possess only a slight CL-PL shift, while others show negligible visible-frequency CL. The first class of emitters exhibits ZPLs predominately around 600-640 nm. 
Certain emitters with ZPLs below 600nm and above 700nm exhibit a $10-15$nm CL-PL spectral shift (i.e., $\mathrm{E_{12}}$, $\mathrm{E_{13}}$, $\mathrm{E_{2}}$, $\mathrm{E_{4}}$, $\mathrm{E_{9}}$, $\mathrm{E_{14}}$, $\mathrm{E_{17}}$, and $\mathrm{E_{18}}$). Possibly, the orbital wavefunction of such emitter energy level is more sensitive to external stimuli, including electron beam induced charging,  carrier screening or the local temperature. Finally, we observe certain emitters with ZPLs in the $650-700$ nm range with negligible CL past 550nm (i.e., $\mathrm{E_{19}}$, $\mathrm{E_{20}}$ and $\mathrm{E_{21}}$ in Fig.~S11). Such a large shift ($>450$ meV) between the PL and CL ZPLs could be due to a change in the emitter charge state~\cite{sontheimer2017photodynamics}, or due to a nearby emitter that is not excited by 532-nm light but captures carriers competitively in the CL process.

\begin{figure}[htbp!]
\centering
    \includegraphics[width=1\textwidth]{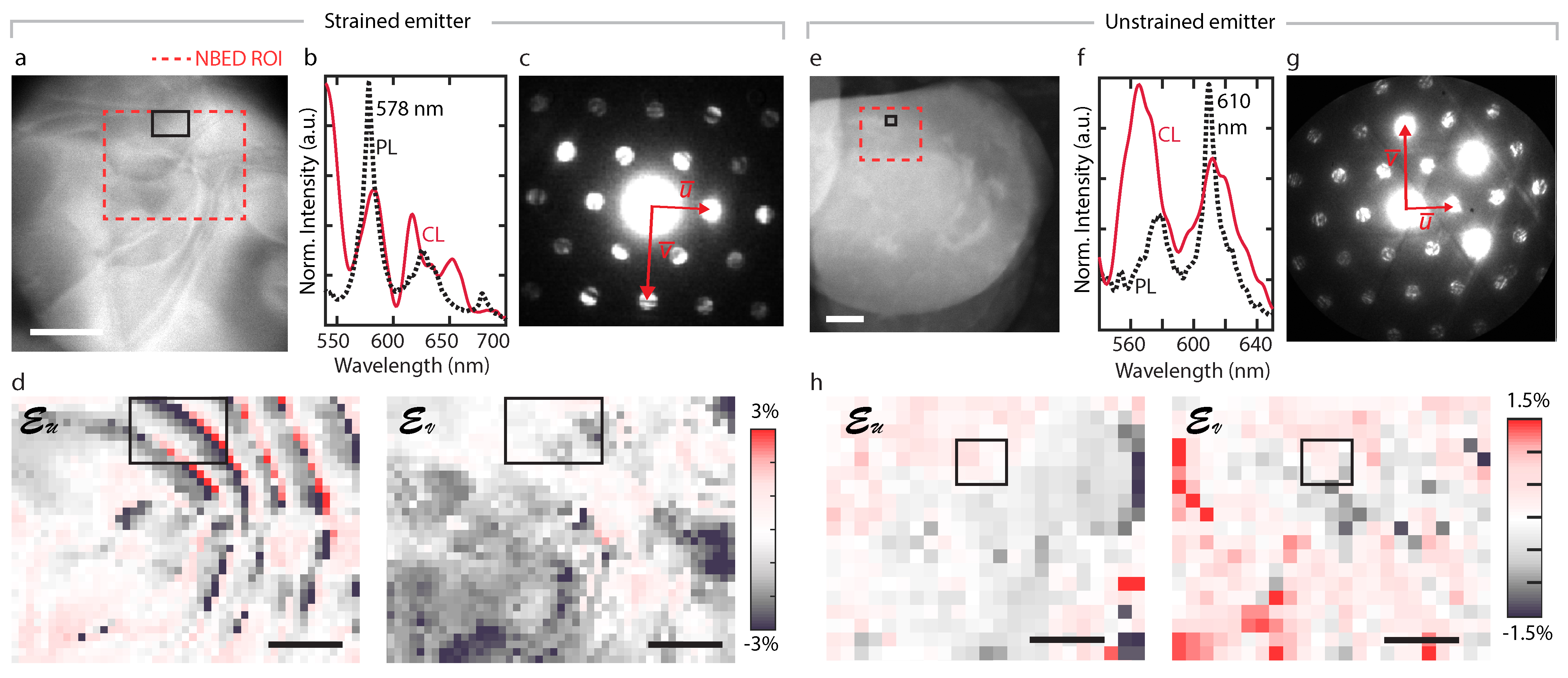}
    \caption{ \textbf{Correlating quantum emission with local strain} (a-b) HAADF image and corresponding PL and CL spectra of one emitter with a 578 nm ZPL. (c) NBED diffraction pattern showing one set of $\vec{u}$ and $\vec{v}$ vectors for calculating strain. (d) Reconstructed strain maps for the NBED ROI region in (a). (e-f) HAADF image and corresponding PL and CL spectra of another region with a emitter at 610 nm. (g-h) Corresponding NBED diffraction image and in-plane strain maps along two directions. Both direction maps show negligible strain in the optically bright pixels. Scale bars are 50 nm for HAADF image and 20 nm for strain maps. The localization of the CL bright spot to the strain map (black rectangle) has at most a 6 nm drift error in both the $x$- and $y$-directions (SI S6).}
    \label{fig:TEMimage}
\end{figure}

We can correlate emitter optical emission with the emitter's local in-plane strain using nano-beam electron diffraction (NBED). We collect diffraction images every 1-3 nm across the emitter hotspot (SI Sec.~S3) and reconstruct in-plane strain maps by fitting the positions of the diffraction disks in each image. The change in lattice distance at each point is calculated by comparing to a reference distance to measure strain. When many flakes stack together, as is common in our samples, we focus exclusively on regions without any lateral rotation between flakes (that is, containing only one set of diffraction disks). Notably, this strain is experienced by the entire stack of hBN flakes. In Fig.~\ref{fig:TEMimage}(a), the emitter $E_{7}$ CL hotspot is marked by a black rectangle in the HAADF image. The emitter's PL and CL can be seen in Fig.~\ref{fig:TEMimage}(b). Diffraction images are collected in the red dotted rectangular region; an averaged diffraction image is shown in (c). The strain maps along $\langle 10\bar{1}0\rangle$ and $\langle \bar{1}2\bar{1}0\rangle$ directions are presented in Fig.~\ref{fig:TEMimage}(d), where the position of the brightest pixel of the CL map is marked with a black rectangle. For this particular emitter, a large strain variation in the CL hotspot is observed compared to the surrounding: from -2.4$\%$ to 2.2 $\%$ along the $10\bar{1}0$ direction and from 0 to -2$\%$ along the $\bar{1}2\bar{1}0$ direction.
On the other hand, as shown in Fig.~\ref{fig:TEMimage}(e-h), we observe the emitter $\mathrm{E_{9}}$ at 610 nm in a relatively unstrained region, indicating strain is not a precursor for emitter formation. To compare strain values across emitters, we fit each region with 2-3 different orthogonal pairs of $\vec{u}$ and $\vec{v}$ vectors and calculate the mean and standard deviations of the CL-hotspot strain values over those fittings.  

\begin{figure}[!htbp]
    \includegraphics[width = 0.5\textwidth, angle = 0]{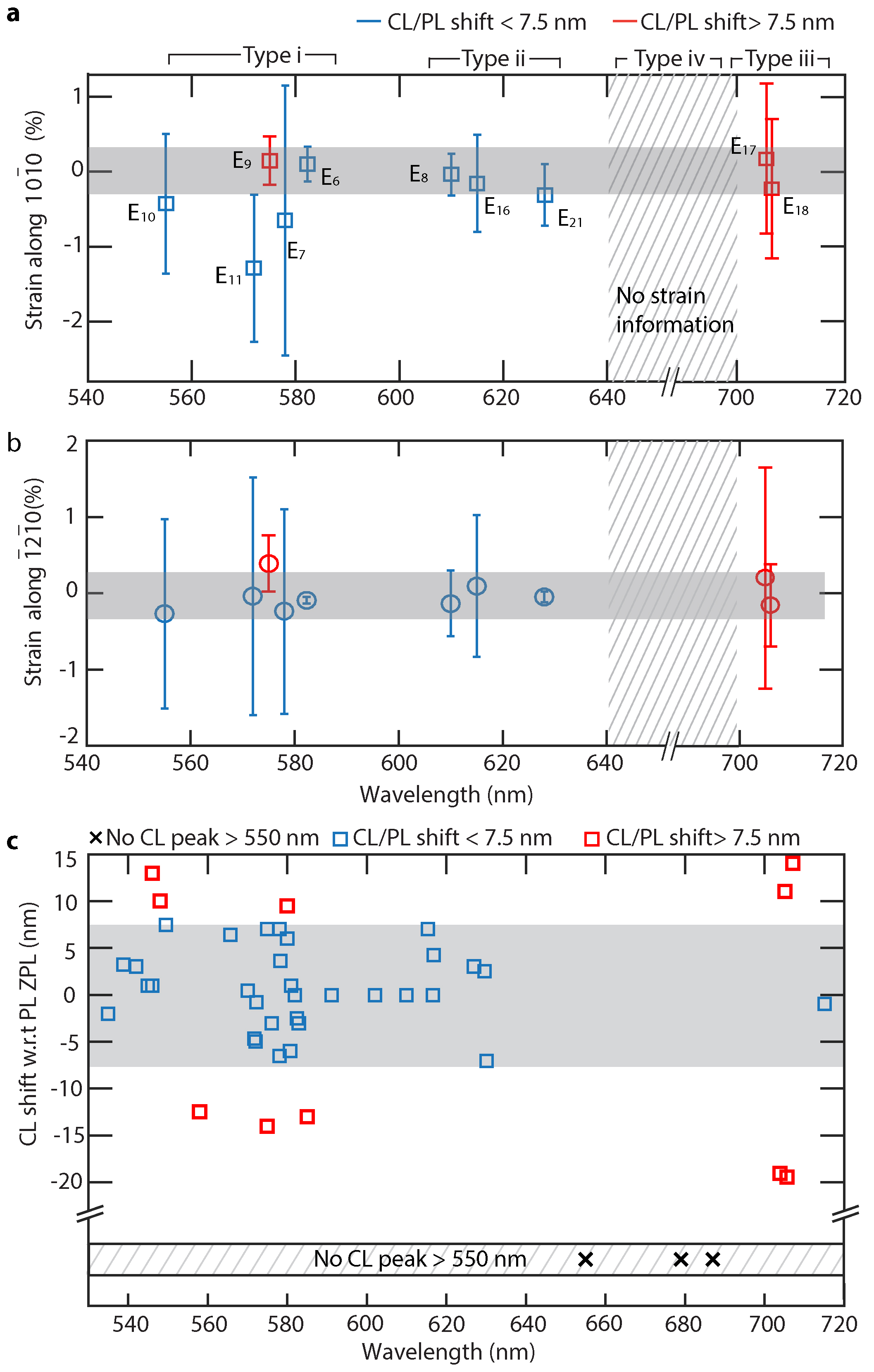} 
    \centering
    \caption{\textbf{Emitter Classification} (a-b) Average strain and standard deviation in the CL hotspot along the $\left< 10\bar{1}0 \right>$ and $\left< \bar{1}2\bar{1}0 \right>$ as a function of emitter emission wavelengths for 10 emitters across 8 different regions of 6 samples. (c) CL spectral shift from the PL ZPL. The grey shaded region corresponds to 1.5x the CL spectrometer resolution; therefore, any emitter within this region is taken to have the same CL-PL ZPL. Black crosses represent emitters that either do not have strong CL peaks or emit broadly in the ultraviolet.}
    \label{fig:strain}
\end{figure}

Fig.~\ref{fig:strain}(a-b) summarizes the mean and standard deviation of strain values in the CL hotspots for 10 emitters along two crystal directions. 
We can distinguish between distinct emitter species from these strain variations combined with their CL-PL responses, summarized in Fig.~\ref{fig:strain}(a-c). Our results indicate that at least 4 classes of emitters exist. First, we observe one type of defect near 580 nm ($\mathrm{E_{11}}$, $\mathrm{E_{7}}$ and $\mathrm{E_{6}}$). These emitters have matched PL-CL spectra but large strain variations among them. The observed spectral tunability (121 meV) between this class of emitters can be due to strain, consistent with previous reports~\cite{grosso2017tunable, xue2018strain} and our first principle modeling (see SI S10). We note that emitter $\mathrm{E_{10}}$ at 554 nm could be part of this same class, even though the strain variation is not more than $1.5\%$ compared to $\mathrm{E_6}$; similar emitters can be seen in Fig.~\ref{fig:strain}(c). We hypothesize that the CL-PL spectral shifts of 10-13 nm ($\sim$ 40 meV) could originate from charging effects due to the electron beam, since an emitter positioned nearer to the carbon substrate will experience a different charging level than one positioned near the middle or top of a hBN flake. 

Next, we see that the three emitters around 615 nm originate from another distinct defect species, which we term Type-\textit{ii}. Notbly, the type \textit{i} emitter $\mathrm{E_{6}}$ (582 nm) and type \textit{ii} emitter $\mathrm{E_{8}}$ (610 nm) are separated by 100 meV with negligible strain difference, making it unlikely that the 615 nm emitters are strain-tuned versions of Type-\textit{i} emitters. 

We identify a third class of emitters around 705 nm. These emitters have shifted CL-PL spectral responses and are shifted from the 630 nm emitter $\mathrm{E_{21}}$ by 200 meV in energy with at most a 1$\%$ strain difference. Such large spectral separation with small strain difference  exceeds theoretically predicated spectral tunability with strain for known defects and defect complexes (see SI Fig.~13 and Ref.~\citenum{grosso2017tunable, xue2018strain}). More examples of such Type-\textit{iii} emitters in panel (c) also show predominantly shifted spectral responses between their PL and CL ZPLs. 

Finally, the emitters around 660 nm have either very low quantum efficiency under electron beam excitation or have a much larger shift ($>450$ meV) in their CL responses ($\mathrm{E_{19}}$, $\mathrm{E_{20}}$ in Fig.~\ref{fig:CL_PL}). Given their very different spectral responses (and hence an uncertaintly in TEM-CL about the precise emitter location), we could not characterize their host region strain. Still, given their largely-shifted PL-CL spectra, we can postulate that they are a distinct defect species (Type \textit{iv}). These results are consistent with Reference ~\citenum{vogl2019atomic}, which hypothesized a distinct defect nature for the 650 nm and the 580 nm emitters.

We can combine our experimental results, first principles modeling, and the community’s understanding of hBN defect structures to interpret our results. Among possible color centers in hBN are boron mono-vacancies ($\mathrm{V_B}$), its hydrogen and oxygen complexes, nitrogen mono-vacancies ($\mathrm{V_N}$), and $\mathrm{C_BV_N}$, $\mathrm{N_BV_N}$ and $\mathrm{O_{2B}V_N}$ (see SI S10 and Reference~\citenum{tawfik:2017}). The calculated ionization energies of B and N in monolayer hBN are 80 kV and 118.6 kV, respectively \cite{kotakoski2010knockondamage}. Therefore, with our electron accelerating voltage of 80kV, we do not expect the mono-boron vacancies ($\mathrm{V_B}$) to be stable. In particular, under 80-120kV electron-beam irradiation, $\mathrm{V_B}$ has only been observed to form in mono- and bi-layer hBN, and once a mono-vacancy is formed, it selectively grows into larger triangles.~\cite{jin2009hBNHRTEM, meyer2009HRTEM, kotakoski2010knockondamage} Similarly, $\mathrm{V_N}$ is not predicted to be a stable single photon emitter and is not active at visible-frequencies. Therefore, our stable defects which have reproducible CL and PL are not these mono-vacancy defects.  

Instead, our Type-\textit{i}-\textit{iv} emitters likely emerge from $\mathrm{C_BV_N}$, $\mathrm{N_BV_N}$ and $\mathrm{O_{2B}V_N}$ and complexes of these defects, as well as $\mathrm{V_N}$ with interstitials (B$_i$, O$_i$, C$_i$).~ \cite{weston2018} For example, $\mathrm{V_N}$ has been postulated to be a donor defect,~\cite{wong2015STM,  yin2010triangle} and a change of its charge state can result in a distinct CL peak from its PL peak; our Type-\textit{iv} defects with no visible-frequency CL but strong SPE emission in PL could be a complex with this donor. In particular, electron beam induced charging can locally modify the potential landscape and hence the defect charge state. This result parallels the disappearance and reappearance of charged defects in bulk hBN due to probe-assisted carrier tunneling, as seen by Wong and colleagues in Ref.~\citenum{wong2015STM}.

In conclusion, we have established a direct correlation between PL, CL, and local crystallographic strain of quantum emitters in hBN. The combined results of these correlated measurements indicated that at least four distinct classes of defects are responsible for quantum emission in hBN. CL hyperspectral mapping reveals multiple emitters within a diffraction-limited optical spot, each contributing to the PL spectra. The large spectral variability of single photon emitters is not due to solely to strain. Additionally, strain is not required to activate the emission, nor are defects are required to be at the edges of the flakes. Importantly, most of the emitters in multilayer films are stable upon electron and optical irradiation - a crucial feature for future atomic-scale imaging. Looking forward, we envision Angstrom-scale imaging of a variety of two-dimensional materials, revealing the rich photonic properties that emerge with atomic-scale architecting and `defects by design'.


\begin{methods}\label{methods}
\textbf{Sample Preparation:} 2.5 $\mu$L of hBN nanoflakes (each flake 1-5 monolayers) in ethanol-water solution (Graphene Supermarket) is dropcast on cleaned holey carbon TEM grid (Quantifoil 0.6/1). The samples are dried in a vacuum bell jar at 80 $^\circ$C for 30 minutes and annealed for 2 hours under ambient pressure and environment at low temperature (200-220$^\circ$C) due to the stability of the ultra-thin carbon TEM substrate. Samples are stored in a glovebox under Ar.\\\\
\textbf{Optical Characterization:} The photoluminescence maps of the hBN nanoflakes are acquired using a Horiba confocal Raman microscope, with a 100x objective (NA = 0.6), an excitation laser power of $\sim$100 $\mu$W at 532 nm, and a grating of 600 lines/mm groove density. The typical laser spot size is $\sim$600nm and the acquisition time is 1 sec for each spectrum. \\\\
\textbf{Autocorrelation g$^{(2)}$  measurements:} Second-order autocorrelation measurements are performed using a Hanbury Brown and Twiss setup. The quantum emitter (QE) is excited by a 532 nm continuous wave laser through a laser line filter, and the photoluminescence from the QE is filtered by a dichroic filter (Semrock FF552-Di02-25x36) and a bandpass filter (bandwidth of 10 nm) centered around the QE emission wavelength, and collected by a multi-mode fiber. The photoluminescence is directed towards a fiber beamsplitter; at the two output ports of the beamsplitter are two single-photon counting modules (SPCMs, Perkin-Elmer SPCM-AQR-14-FC). One SPCM is used as a start signal and the second used as a stop signal. By measuring the time delay between the two successive photon arrivals, a histogram of occurrences as a function of time delay is constructed. The photon counts were correlated using a PicoHarp300 time-correlated single-photon counting module, with a count rate of ~10k cps and a typical integration time of 20 minutes for one dataset. For this measurement, the excitation laser power is kept below 20$\mu$W to prevent degradation of both the emitters and the thin carbon film ($\sim$12 nm) TEM grid. The counts are normalized by counts at 4 $\mu$s and fitted assuming three level system (more in SI S2). \\\\
\textbf{STEM-Cathodoluminescence characterization:} The electron microscope FEI Titan is operated at 80 kV. The Gatan Vulcan Cathodoluminescence TEM holder is used for light collection in STEM. The holder has two mirrors positioned above and below the sample. A hole of about 500 $\mu$m in the mirrors allows the electron beam through. The reflected light is focused onto two optical fibers (multimode step index) and directed onto the photomultiplier tube (PMT) for panchromatic mapping or to the grating and CCD (cooled to -60$^\circ$C) for spectral mapping. The two mirrors in the CL holder are focused and the collection window is 200 $\mu$m in diameter. We use two fiber-coupled (multimode) LEDs (red and green) to align the two hotspots from the two mirrors after loading the sample. \\
The panchromatic photon pulse counting maps using the PMT are collected as 512 x 512 pixels and each pixel acquisition time is 60/100 $\mu$s. For this mapping, C2 aperture is chosen to be 70 $\mu$m (convergence semi angle is 14 mrad) and the STEM spot size (non monochromated) is chosen as either 6 or 7 (depending on the emitter intensity) and the screen current is 20 pA (spot size 7) or 40 pA (spot size 6). The spectral maps are collected using a 150 lines$/$mm grating and an acquisition times of 4s$-$8s per pixel, again depending on the intensity of the particular emitter. The spot size for this mapping is 6 and screen current is 40 pA. Software drift correction (after every 3 rows) is on during the collection. Samples are usually very stable with minimal drift. Dark corrections are performed at the end of collection of each maps. All mapping is done at room temperature.\\\\
\textbf{Correlation of optical and electron micrographs:} The TEM substrate is supported by a Au grid where each square in the grid is 40 $\mu$m x 40 $\mu$m. We focus on the central 4 squares, as those can be positioned to be in the CL holder optical hotspot ($\sim 200 \mu$m in diameter). The 4 corners of each square serve as alignment markers for the region of interest; we note coordinates of each corner and each emitter region to triangulate distances from each corner. We take optical images of the entire square with a 100X objective and mark these regions with respect to the corners. Before STEM-CL imaging, we obtain TEM images at 70X-100X magnification (LM mode). We then use previously calculated distances from the corners and the 100X optical image to locate the regions of interest; this method is accurate to 1-2 $\mu$m. For higher-accuracy mapping, we use a numbering system for the holes (i.e., recording the number of holes in $x$ and $y$ direction from any specific corner) to determine the appropriate position in the TEM. For localization beyond the diffraction limit, we collect STEM panchromatic-CL maps as described in Fig.~\ref{fig:experimentation}(b). This method allows us to determine how many `bright' emitters are located inside the $\sim$500 nm optical PL hotspot. Next, we collect spectral signatures of each bright region as described in main text. \\\\
\textbf{NBED diffraction experiment:}  The C2 aperture size is chosen as 10 $\mu$m (convergence semi angle of 2 mrad), the spot size is 11, the screen current is $\sim$1 pA, each pixel acquisition time is 0.1 s, and each diffraction image is 4k x 4k. \\\\ 
\textbf{CL Spectral decomposition:} Spectral decomposition is performed using HyperSpy package in Python \cite{hyperspy}. Any noise peaks are removed and the spectra are smoothed with LOWESS/ LOESS weighted regression algorithm using a smoothing parameter of 0.04. We perform principal component analysis (PCA) on the dataset to estimate the number of spectral components from the PCA weights. Finally, non-negative matrix factorization is performed with this number of spectral components to calculate the spectral weights.\\\\
\end{methods}


\bibliography{sample}
\begin{addendum}
 \item The authors acknowledge useful discussions about strain mapping with Dr.~Thomas Pekin (Humboldt-Universit\"at zu Berlin) and about CL-PL correlations with Dr.~Claire McLellan (Stanford). TEM imaging and spectroscopy were performed at the Stanford Nano Shared Facilities (SNSF) and Stanford Soft \& Hybrid Materials Facility (SMF). The PL, TEM experiments and theoretical modeling were supported by the DOE `Photonics at Thermodynamic Limits' Energy Frontier Research Center under grant DE-SC0019140. F.~H.~gratefully acknowledges the support of the Diversifying Academia, Recruiting Excellence (DARE) Doctoral Fellowship Program by Stanford University. J.~V.~acknowledges the support the NSF Quantum Leap  EAGER Grant number DMR 1838380. T~.F.~H. and L.~Y.~also acknowledge support from the Betty and Gordon Moore Foundation EPiQS Initiative through grant no.~GBMF4545.
 \item[Competing Interests] The authors declare that they have no competing financial interests.
 \item [Author Contributions] F.~H.~, L.~Y.~, T.~F.~H.~ and J.~A.~D.~conceived the project. F.~H.~made the samples, performed the PL and TEM experiments; J.~L.~Z.~, L.~Y.~ and F.~H.~performed the second-order correlation measurements under the supervision of J.~V.~; C.~C.~ and P.~N.~performed the theoretical calculations. A.~F.~M.~assisted in strain map data collection. M.~N.~and I.~A.~assisted in sample preparation. F.~H.~and J.~A.~D.~wrote the draft. J.~A.~D.~supervised the project. All  coauthors discussed the results and provided useful feedback on the manuscript. 
 \item[Correspondence] Correspondence and requests for materials should be addressed to F.~H.~and J.~A.~D. 
\end{addendum}

\end{document}